\documentstyle[aps,prb,eqsecnum,psfig,floats,twocolumn]{revtex}

\draft
\begin{document}

\pagestyle{empty}

\twocolumn[\hsize\textwidth\columnwidth\hsize\csname@twocolumnfalse%
\endcsname

\title{Nonlinear $\sigma$ Model Treatment of Quantum Antiferromagnets 
in a Magnetic Field}

\author{B. Normand, Jordan Kyriakidis, and Daniel Loss}

\address{Departement f\"ur Physik und Astronomie, Universit\"at Basel,
Klingelbergstrasse 82, CH-4056 Basel, Switzerland. }

\date{\today}

\maketitle

\begin{abstract}

We present a theoretical analysis of the properties of low-dimensional 
quantum antiferromagnets in applied magnetic fields. In a nonlinear 
$\sigma$ model description, we use a spin stiffness analysis, a 1/N 
expansion, and a renormalization group approach to describe the 
broken-symmetry regimes of finite magnetization, and, in cases of 
most interest, a low-field regime where symmetry is restored by 
quantum fluctuations. We compute the magnetization, critical fields, 
spin correlation functions, and decay exponents accessible by nuclear 
magnetic resonance experiments. The model is relevant to many systems 
exhibiting Haldane physics, and provides good agreement with data for 
the two-chain spin ladder compound CuHpCl.

\end{abstract}

]

\section{Introduction}

The importance of low-dimensional spin systems in revealing
fundamental quantum mechanical properties has been recognized since
Haldane's conjecture\cite{rh} concerning the effects of quantum
fluctuations in integral- and half-integral-spin antiferromagnetic
(AF) chains. The key quantity is the topological term arising from
the quantum spin phase, and similar considerations have since been
extended to anisotropic and higher-order interactions, higher
dimensions (planes), and coupled chains (spin ladders), largely
driven by materials and experiment. Magnetic flux within a sample
can alter the effective spin magnitude, and features visible in an
applied field include spin gaps and magnetization plateaus.\cite{roya} 

Recent progress in metalloorganic synthesis has furnished new  
classes of low-dimensional antiferromagnets (AF) whose small 
exchange constants make their full magnetization response, including 
saturation, accessible to laboratory fields. The measured magnetization 
curves of some such materials, thought to be prototypical quantum 
magnets, in fact exhibit certain features which are rather classical. 
Examples on which experimental interest has focused are the 
Haldane ($S$ = 1) chain NENP,\cite{ragksi} the planar
system CFTD,\cite{rhcmmv} and the two-chain spin ladder 
CuHpCl.\cite{rcclpmm,rhrbt} We will focus on the ladder geometry 
of CuHpCl, the best-characterized sample in recent literature.

We consider the quantum AF system in an external magnetic field
using the nonlinear $\sigma$ model (NLsM). While this widely-applied
treatment is in fact semiclassical, being truly valid only in the
limit of large on-site spin $S$, it has in the past formed the basis
for many fundamental deductions concerning the quantum limit of
antiferromagnetic spin systems\cite{ria,rchn,rem}. We will demonstrate 
here its validity in the case of all effectively integral-spin quantum 
systems in appreciable magnetic fields, and provide justification 
for this result in terms of suppression of quantum fluctuations by 
the field. Such systems display an ideal quantum phase transition,
driven by the applied field, between a disordered regime with 
gapped spin excitions at low field and a quasi-long-range ordered 
regime with gapless excitations at higher field. 

This model presents an ideal example of symmetry breaking in 
condensed matter systems. The O(3) symmetry of the spin Hamiltonian, 
and of its low-energy description in terms of the NLsM, is broken on 
a purely classical level. However, at low fields this symmetry is 
completely restored by strong quantum fluctuations, a well-known 
property of the NLsM, which has been used to illustrate the effect of 
asymptotic freedom (or ``confinement of excitations").\cite{rp}
In contrast, quantum fluctuations may be suppressed by fields which 
are sufficiently strong but remain experimentally accessible, with the
consequence that the full O(3) symmetry cannot be restored, and the
spin system is reduced to the lower XY or O(2) symmetry. This XY 
symmetry, in the plane perpendicular to the applied field direction, 
remains sufficient that there can be no true long range order, 
but instead a quasi-long-range order characterized by spin-spin 
correlation functions which decay in space-time with a power law form. 

The outline of this paper is as follows. In Sec.~II we derive the 
form of the NLsM in a magnetic field, using the geometry of the 
two-chain ladder system. With a view to experimental comparison we 
include magnetic saturation by applying a total-spin constraint. 
In Sec.~III we analyze the spin stiffness of the model, which allows 
us to deduce the general behavior, correlation length, spin gap and 
critical field at zero temperature, and also the effects of finite 
temperature and system size. In Sec.~IV we consider the $1/N$ 
expansion, which is well suited for describing the low-field, 
disordered regime. Sec.~V contains an extensive renormalization group 
(RG) study of the model over the entire field range. We derive the 
coupled RG equations for the 1+1-dimensional (1+1d) system and 
present their general solution. This allows us to show how the 
equations return the physics of symmetry-breaking and -restoration 
in the previous paragraph, and to discuss the issue of 
renormalization and gauge-invariance. We compute in Sec.~VI the 
magnetization of the model using a high-field expansion, and in 
Sec.~VII calculate spin correlation functions for the low- and 
high-field regimes. In Sec.~VIII we compare our results with 
experimental data from magnetization and nuclear magnetic 
resonance (NMR) spin-relaxation measurements. Sec.~IX summarizes 
our conclusions, and discusses the variety of systems and physical 
problems to which the formalism is applicable. 

\section{Nonlinear $\sigma$ Model}

\begin{figure}
\centerline{\psfig{figure=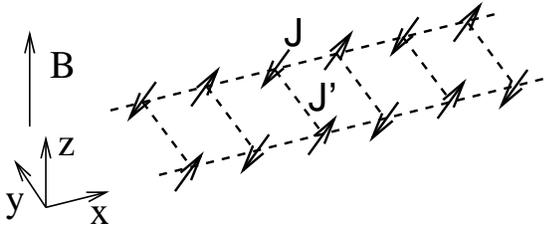,height=3.0cm,angle=0}}
\medskip
\caption{Representation of two-chain ladder system and coordinate 
axes. $J = J_x$ is the intrachain and $J^{\prime} = 2 J_y$ the 
interchain exchange coupling. }
\end{figure}

We consider the ladder geometry of the compound CuHpCl.\cite{rcclpmm} 
The Hamiltonian for the system in a magnetic field
${\bf b} =
\tilde{g} \mu_B {\bf B}$ may be written as
\begin{equation}
   {\hat H} = \sum_{i; m=1,2}
   \left(
      J {\hat {\bf S}}_{m,i} \cdot {\hat {\bf S}}_{m,i+1} +
      J^{\prime} {\hat {\bf S}}_{1,i} \cdot {\hat {\bf S}}_{2,i} +
      {\bf b} \cdot {\hat {\bf S}}_{m,i}
   \right),
\label{esh}
\end{equation}
where $J$ is the superexchange interaction between spins in each
chain and $J^{\prime}$ the interchain, or ladder ``rung'',
interaction. We choose $N_x$-site chains with periodic boundary
conditions in both directions, and the geometry shown in Fig.~1,
such that $J_x = J$, $J_y = J^{\prime} / 2$, ${\bf B} = (0,0,B)$,
and ${\hat {\bf S}}_{m,N_x + i} = {\hat {\bf S}}_{m,i}$. Following 
the standard procedure in the derivation of the 
NLsM,\cite{rh,rmam,ref,raa,rl} the coherent-state representation 
of the spin ${\hat {\bf S}}_{m,i}$ is written as $S {\bf 
\Omega}_{m,i} \simeq S[(-1)^{i+m} {\bf n}_{m,i} + a {\bf 
l}_{m,i}]$, where ${\bf n}_{m,i}$ is a staggered spin (the N\'eel 
vector) with unit magnitude, and ${\bf l}_{m,i}$ describes the spin 
fluctuations perpendicular to ${\bf n}_{m,i}$.  On proceeding to the 
continuum limit, the Hamiltonian is expressed in coherent-state 
representation as
\begin{equation}
   H = \int\!d {\bf r}
   \left(
      \frac{1}{2} S^2 \sum_{j = x,y} J_{j}
      \left[
         \left( \partial_{j} {\bf n} \right)^2 + 4 {\bf l}^2
      \right] -
      \frac{S}{a} {\bf l} \cdot {\bf b}
   \right).
\label{efnlsm}
\end{equation}

The full Euclidean action ${\cal S}_E = {\cal S}_{WZ} + 
\int_{0}^{\beta} d \tau H$ in space ${\bf r} = (x,y)$ and inverse 
temperature $\tau$ ($\beta = 1 / k_B T$) contains a Berry-phase, 
or Wess-Zumino, term whose origin lies in the solid angle subtended 
by the fluctuating quantum spin phases:
\begin{equation}
   {\cal S}_{WZ} = i \frac{S}{a^2} \int\! d\tau \, d{\bf r} \,
   \left[
      \dot{\phi} - a {\bf l} \cdot
      \left( {\bf n} \wedge \dot{\bf n} \right)
   \right]
   + 4 \pi i S \left( P_1 + P_2 \right).
\label{eswz}
\end{equation}
Here, $\phi$ is the azimuthal angle, and $P_1 = P_2 = (1 / 4\pi)
\int d\tau \, dx \, ({\bf n} \wedge \dot{\bf n}) \cdot \partial_x
{\bf n}$ are topological terms which give the Pontryagin index on
each chain when $\partial_y {\bf n} = 0$, as is the case in a ladder
of only two chains.\cite{gf} The last term in Eq.~(\ref{eswz}) is 
therefore $i(4 \pi P_1) 2 S$, demonstrating that the system will have
integral-spin characteristics for any value of $S$, and the
topological term may thus be ignored.\cite{rk,rs}

Integrating out the fluctuations ${\bf l}$ about the staggered spin
configuration of ${\bf n}$, subject to the orthogonality constraint
${\bf n} \cdot {\bf l} = 0$, yields the classical solution\cite{rl,ral}
\begin{equation}
   {\bf l}_{\text{cl}} = \frac{1}{4 a S \bar{J}} \left\{
      i \left( {\bf n} \wedge \dot{\bf n} \right) + \left[
         {\bf b} - {\bf n} \left( {\bf b} \cdot {\bf n} \right)
      \right] \right\},
\label{elcl}
\end{equation}
where $\bar{J} = J_x + J_y$.  Finally, the corresponding action for
the quasi-one-dimensional (1D) ladder system, in 1+1 Euclidean
dimensions denoted by $\mu$, is
\begin{eqnarray}
   {\cal S}_E = \frac{1}{2 g} \int\! d\tau \, dx \,
   \left\{
      \left( \partial_{\mu} {\bf n} \right)^2 -
      \left[
         {\bf b}^2 - \left( {\bf n} \cdot {\bf b} \right)^2
      \right] \frac{}{}
   \right. \nonumber \\
   \left. \frac{}{}
      \mbox{} +
      2i {\bf b} \cdot \left( {\bf n} \wedge \dot{\bf n} \right)
   \right\},
\label{ese}
\end{eqnarray}
where $g = (2 / N_y S) \sqrt{\bar{J} / J_x}$ is the bare coupling
constant, and the integral over $\tau$ is from zero to $L_T = c
\beta$, with $c = 2 S a \sqrt{J_x \bar{J}}$ the effective spin-wave
velocity. (We use $\hbar = 1$ throughout.)  Summation over $\mu$ is 
implied in the square. We have left explicit the number of chains 
$N_y$ in the ladder.  For CuHpCl we have that $J_y > J_x$ and $N_y 
= 2$, whence the rung coupling may be treated as effectively rigid.
The form of the NLsM in an external field given by Eq.~(\ref{ese})
has been derived previously,\cite{rmam} and its implications
considered in the low-field limit.\cite{ral}  We will examine in 
this work its properties at all field strengths. 

The action (\ref{ese}) may be cast in a physically more transparent 
form by expressing the staggered spin in spherical coordinates, 
${\bf n} = (\sin\theta \cos\phi, \sin\theta \sin\phi, \cos\theta)$. 
From the resulting Lagrangean density
\begin{equation}
{\cal L}_{\rm E} = \frac{1}{2 g} \left\{ \left[ \left( \dot{\phi} 
+ i b \right)^2 + \left( \partial_x \phi \right)^2 \right] \sin^2 
\theta + (\partial_{\mu} \theta)^2  \right\}
\label{eldep}
\end{equation}
one observes that the field induces a hard-axis anisotropy through 
the term $({\bf n} \cdot {\bf b})^2$, which makes the spins tend to 
align with the N\'eel vector ${\bf n}$ in the plane normal to ${\bf 
b}$. For strong fields, deviation of ${\bf n}$ from this plane is 
small, and expansion to quadratic order in $\vartheta = \theta - 
\pi/2$ is valid. Taking the field along the $z$-axis, we find that 
the Lagrangean 
\begin{eqnarray}
   {\cal L}_{\rm E} = \frac{1}{2 g}
   \left\{
      \left[ \left( \dot{\phi} + i b \right)^2 +
         \left( \partial_x \phi \right)^2
      \right]
   \right. \nonumber \\
   \left.
      \mbox{} - \vartheta
      \left[
         \partial_{\mu}^2 + \left( \dot{\phi} + i b \right)^2
      \right] \vartheta
   \right\}
\label{elde}
\end{eqnarray}
separates into contributions from in-plane and out-of-plane
fluctuations in the high-field regime, $|b| > \dot{\phi} \equiv 
\omega$. We will use Eq.~(\ref{elde}), and return to the condition 
for its validity, below.

Finally, for a real magnetic system with small spin quantum 
number $S$, application of a sufficiently large magnetic 
field will cause total spin alignment, or saturation, with a 
maximum magnetization given by $M_s = \tilde{g} \mu_B S N_s$, 
where $N_s$ (= $N_x N_y$ here) is the total number of sites in 
the system. This effect is not contained in the NLsM 
(Eq.~(\ref{ese})) discussed above, where $S$ is required to be 
large (semi-classical). Furthermore a saturation of the spin would 
correspond to large $|{\bf l}|$ and small $|{\bf n}|$, rendering 
invalid the expansion of ${\bf S}$ (below Eq.~(\ref{esh})). We will 
include saturation by placing a limit, corresponding to the desired 
experimental spin magnitude, on the spin component ($S_z$) projected 
along the field. By this technique, the full spin vector remains 
primarily in the plane, justifying the treatment of the uniform spin 
component $|{\bf l}|$ as a small variable. 

The succeeding experimental comparison (Sec.~VIII) will be simplified 
by treating $M_s$ as a simple cutoff, but we describe here briefly 
this systematic inclusion of saturation in the model by the 
application of a constraint on the total spin. In addition to the 
constraints ${\bf n \cdot l} = 0$ on the uniform and staggered 
spin components, and ${\bf n}^2 = 1$ on the magnitude of the 
staggered spin, we introduce a further constraint enforcing 
conservation of spin. For a field applied along the $z$ axis, 
the conserved quantity is the $z$-component of the total spin 
$S^z \equiv \langle \{{\bf \Omega}\} | \widehat{S}^z | \{{\bf 
\Omega}\} \rangle \rightarrow S\int\!dx\, l^z$, where the continuum 
limit is taken in the final step.  While the previous constraints 
are purely local, in that they must be enforced at each spacetime 
point, the spin constraint is local in time but global in space. All 
of the constraints are treated by inserting (functional) delta 
functions into the measure of the propagator, in the form
\begin{eqnarray}
   \lefteqn{ \left\langle \left\{{\bf \Omega}\right\}_b \right|
      e^{-\beta H} \left| \left\{{\bf \Omega}\right\}_a
      \right\rangle }
   \ \ \ \ \ \ \nonumber \\ && =
   \int\! {\cal D}{\bf n}(x,\tau)
   \!\int\! {\cal D} {\bf l}(x,\tau)\,
   e^{- f_{\rm C} [{\bf n}, {\bf l}]} e^{-{\cal S}_E[{\bf n}, {\bf l}]} ,
\end{eqnarray}
where the function $f_{\rm C} [{\bf n}, {\bf l}]$ specifies the 
constraints using Lagrange multipliers. Integrating over the fluctuating 
spin field ${\bf l}$, which is now constrained, and over all the Lagrange 
multipliers, one obtains the Euclidean action
\begin{eqnarray}
   {\cal S} [{\bf n}] &=& {\cal S}_{WZ} +
   \frac{1}{2g} \int d\tau\,dx \,
             \left( \partial_{\mu} {\bf n} \right)^2 
   \nonumber \\ && \mbox{} -
   \frac{1}{2g} \int d\tau\,
   \frac{\left(D[{\bf n}] + 2iN_s\sigma\right)^2}{A[{\bf n}]} -
   S N_s \beta b \sigma,
   \label{eq-c.con}
\end{eqnarray}
where $D[{\bf n}] = \int\! dx\, ({\bf n} \wedge \dot{{\bf n}})_z$
and $A[{\bf n}] = \int\! dx\, (1 - n_z^2)$, and we have introduced the
scaled spin $\sigma = S_z / (S N_s)$, $-1 \le \sigma \le 1$. In this
action the field $b$ is coupled only linearly to the spin $\sigma$, 
as in ferromagnetic systems, and in contrast to Eq.~(\ref{ese}), 
where it appears quadratically. This linear coupling is 
a direct consequence of the spin constraint, and one may 
recover the previous action (Eq.~(\ref{ese})) by allowing $\sigma$ to 
fluctuate in time, relaxing the condition $|\sigma| \le 1$, and 
integrating over $\sigma (\tau)$. Evaluation of the partition function 
with saturation, requires summing the constrained action 
(Eq.~(\ref{eq-c.con})) over all degrees of freedom, which include the 
allowed values of $|S_z| \le N_s S$.  This procedure limits $M$
to $M_s$ at large fields, and provides a consistent description of 
a second phase transition, to saturation, which occurs at high fields. 
In Secs.~III-V we will concentrate on the symmetry-breaking quantum 
phase transition (Sec.~I), which occurs at a lower field on the 
order of the zero-field gap (below), and so the discussion there 
will not involve further consideration of the spin constraint. 

\begin{figure}
\centerline{\psfig{figure=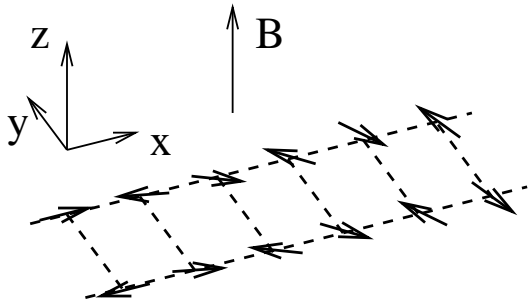,height=4.0cm,angle=270}}
\medskip
\caption{Ladder system with twisted staggered moment.}
\end{figure}

\section{Spin Stiffness}

To gain initial insight into the effect of the magnetic field, we
consider the spin stiffness of the ladder system using the method of
Ref.\onlinecite{rlm}.\cite{rzj} We take (Fig. 2) the 
staggered spin configuration to be subject to a twist $\psi$ in the 
plane, ${\bf n} (\tau,0) = (1,0,0)$ and ${\bf n} (\tau,L) = (\cos 
\psi, \sin \psi, 0)$. This is conveniently represented by ${\bf n} = 
{\cal R}(x)${\boldmath $\sigma$}, where ${\cal R}$ is a matrix for 
rotation by $\psi x / L$ about the $z$ axis, and {\boldmath $\sigma$} 
= {\boldmath $\sigma$}$(\tau,x)$ a 3-component vector with value 
$(1,0,0)$ at the spatial endpoints. The cross-term in $(\partial_{\mu} 
{\bf n})^2$ and the third term in Eq.~(\ref{ese}) are total 
derivatives in $x$ and $\tau$ respectively, and will be discarded 
for the present purposes. This is equivalent to considering only 
paths with no winding in Euclidean space, and gives the action the 
form
\begin{equation}
   {\cal S}_E = \frac{1}{2 g} \int\! d \tau \, dx  \,
   \left\{
   \left( \partial_{\mu} \mbox{\boldmath $\sigma$} \right)^2 +
   \left( \frac{\psi^2}{L^2} - \frac{b^2}{c^2} \right)
   \left( 1 - \sigma_{3}^2 \right) \right\} .
\label{esss}
\end{equation}
To 1-loop order in $g$, the spin stiffness is given by
\begin{eqnarray}
\rho_s & = & \frac{1}{2} c L \left. \frac{\partial^2 F}{\partial \psi^2}
\right|_{\psi = 0} \label{ess1l} \nonumber \\ & = & \frac{c L}{2 L_T}
\left. \frac{\partial^2} {\partial \psi^2} \right|_{\psi = 0} \left\{ 
\frac{L L_T}{2 g} \left[\frac{\psi^2}{L^2} - \frac{b^2}{c^2} \right] 
\right. \nonumber \\ & & \;\;\;\; + \left. \frac{1}{2} {\rm Tr} \ln
\left[{\bf k}^2 - \frac{\psi^2}{L^2} + \frac{b^2}{c^2} \right] \right\} 
\nonumber \\ & = & \rho_s^0 \left[ 1 - \frac{g}{L L_T} \sum_{\bf k} 
\frac{1}{{\bf k}^2 + (b/c)^2} \right],
\end{eqnarray}
where $\rho_s^0 = c/2g$ is the classical (bare) value, and the sum 
includes both quantum and thermal (through the finite ``length'' $L_T$)
corrections to first order in $g$. 

We consider first the low-temperature, or ``quantum'' regime $L_T \gg 
L$. We evaluate the summation between the spatial limits $\pi/L$ and 
$\pi/a$ (this change to open boundary conditions does not affect the 
result), and introduce the ``magnetic length'' 
\begin{equation}
L_m = \pi c / b 
\label{eml}
\end{equation}
to obtain the expression
\begin{equation}
\rho_s = \rho_s^0 \left[ 1 + \frac{g}{2 \pi} \ln \left( \frac{a/L +
\sqrt{(a/L)^2 + (a/L_m)^2}}{( 1 + \sqrt{1 + (a/L_m)^2)}} \right)
\right].
\label{efsse}
\end{equation}
The system length $L$ in Eq.~(\ref{efsse}) may be substituted by a
correlation length $\xi$, beyond which segments of the ladder behave
independently, and $\xi$ then computed from the condition $\rho_s = 0$.
We emphasize that $\rho_s$ is to be considered as a local stiffness,
meaningful only on length scales $L \le \xi$. In the zero-field
limit ($L_m \rightarrow \infty$) one obtains the result $\xi_0 = 
A a e^{2 \pi / g} = A a e^{\alpha \pi S}$, where $\alpha = \sqrt{J_x 
/ \bar{J}}$ introduces a dependence on the ladder coupling 
constants,\cite{rs} and $A$ is a nonuniversal constant of order unity 
which depends on the integration cutoff. The general solution can be 
expressed as
\begin{eqnarray}
\xi(B) & = & 2 L_{m}^{*} / [ 1 - (L_{m}^* / L_m )^2 ] \label{exib}
\nonumber \\ & \equiv & \xi_0 / [ 1 - (B / B^*)^2 ],
\end{eqnarray}
from which we see that the correlation length $\xi(B)$ diverges at 
a critical field $B^*$, where $L_{m}^* = a \sinh(2 \pi / g)$. For
fields $B < B^*$, the system has only short-range correlations, 
and the finite correlation length may be written as $\xi(B) = a
e^{\alpha \pi \tilde{S}}$, where $\tilde{S} = S \left[ 1 - (g / 2
\pi) \ln \left( 1 - ( L_{m}^* / L_m )^2 \right) \right]$ is a
growing value of the effective spin. There is no spontaneous
breaking of the $O(3)$ spin symmetry. For $B > B^*$, the field
enforces a long-ranged correlation throughout the system,\cite{ra} 
and it is most convenient to write the spin stiffness as $\rho_s =
\rho_{s}^0 \left[ 1 - (g / 2 \pi) \ln ( 1 + L_m / a ) \right]$, a
quantity which recovers the bare value $\rho_{s}^0$ as $B 
\rightarrow \infty$. Finally, at the transition $B^*$, it is clear 
from Eq.~(\ref{exib}) that the diverging correlation length 
corresponds to the closing of a gap $\Delta \sim \xi^{-1}$ to spin 
excitations with the form 
\begin{equation}
\Delta \propto 1 - \left( B / B^* \right)^2. 
\label{egfd}
\end{equation}
This situation is summarized in Fig. 3. While $\Delta$ is the gap 
to all fluctuations in the low-field regime where $O(3)$ spin 
symmetry is maintained, we note that at high fields not all the 
modes are gapless. In this regime (strictly $B \gg B^*$) the 
symmetry is lowered to $O(2)$, and while in-plane ($\phi$) 
fluctuations (Eq.~(\ref{elde})) are massless, excitation modes 
($\vartheta$) in the direction of the field, or out of the plane 
it enforces, are gapped with ``mass'' $b$. 

\begin{figure}
\centerline{\psfig{figure=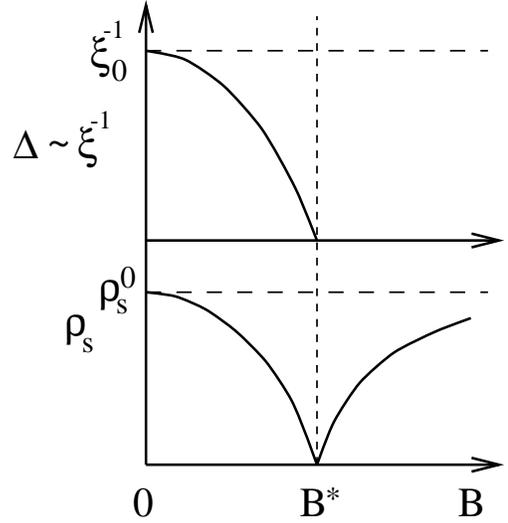,height=7.0cm,angle=270}}
\medskip
\caption{Schematic behavior of spin stiffness and correlation
length, or spin gap, with applied field. }
\end{figure}

We may further employ the spin stiffness as a means of characterizing 
the behavior of a system of finite size $L$, and at finite temperatures, 
for different values of the magnetic field. This is essentially a 
matter of comparing the corresponding length scales $L, L_T$ and 
$L_m$\cite{rlm}. We present results for the cases of i) weak and ii) 
strong fields in the following three regimes: a) {\it quantum}, 
i) $L_m, L_T \gg L$ and ii) $L_T \gg L \gg L_m$, b) {\it classical 
renormalized}, i) $L_m, L \gg L_T$ and ii) $L \gg L_T \gg L_m$, and 
c) {\it classical}, i) $L_m, L \gg a \gg L_T$ and ii) $L \gg L_m, 
a \gg L_T$. Following Ref.\onlinecite{rlm} we quote the stiffness 
in dimensionless units of $\rho_s / c = (1 / 2 g) (\rho_s / \rho_{s}^0)$. 
For weak fields we find 
\begin{equation}
\frac{\rho_s}{c} = \left\{ \begin{array}{ll} 
\frac{1}{4 \pi} {\rm ln} \left[ \frac{\xi_o}{2} \left( \frac{1}{L} + 
\sqrt{\frac{1}{L^2} + \frac{1}{L_{m}^2}} \right) \right], \; & L_m, L_T 
\gg L \\ ( \xi_B (T) - {\rm min} [L,L_m]) / 12 L_T , & L_m, L \gg L_T 
\\ ( \xi_{\rm cl} - L) / 12 L_T , & L_m, L \gg a \gg L_T . \end{array}
\right.
\label{ewfss}
\end{equation}
$\xi_0$ denotes the quantum correlation length defined above, 
$\xi_{\rm cl} = 6 L_T / g$ is the classical correlation length, and 
\begin{equation}
\xi_B (T) = \frac{3 L_T}{\pi} \left[ {\rm ln} \left( C \frac{\xi_0}
{2L_T} \sqrt{1 + \sqrt{1 + (L_T / L_m)^2}} \right) \right] ,
\label{ewfrccl}
\end{equation}
with $C \sim O(1)$ another nonuniversal constant, contains the 
renormalizing effect of quantum fluctuations on $\xi_{\rm cl}$. 
These results are all directly analogous to the zero-field 
case,\cite{rlm} with the field entering as a quadratic correction 
where relevant. In a strong applied field (ii), 
\begin{equation}
\frac{\rho_s}{c} = \left\{ \begin{array}{ll} 
\frac{1}{4 \pi} {\rm ln} \left[ \frac{\xi_o ( \sqrt{ 1 + (L_m / L)^2})}
{L_m + \sqrt{ a^2 + L_{m}^2 }} \right], \; & L_T \gg L \gg L_m \\
( \xi_T (B) - L_m ) / 12 L_T , & L \gg L_T \gg L_m \\ 
( \xi_{\rm cl} - L_m ) / 12 L_T , & L \gg L_m, a \gg L_T , \end{array}
\right.
\label{esfss}
\end{equation}
where now the renormalized classical correlation length has the 
field-dominated form 
\begin{equation}
\xi_T (B) = \frac{3 L_T}{\pi} \left[ {\rm ln} \left( C \frac{\xi_0}
{L_m} \sqrt{\sqrt{1 + (L_m / L_T)^2} + L_m / L_T} \right) \right] .
\label{esfrccl}
\end{equation}
The general characteristic of all the above results is a logarithmic 
dependence of the spin stiffness on the length scales of system size, 
temperature, and field where quantum fluctuations are important, 
turning to a power-law dependence in the classical regime. In the 
high-field situation the scale $L_m$ of the field replaces $L_T$ 
as the determining parameter ({\it cf.} Eqs. (\ref{ewfss},
\ref{esfss})). Equivalent results for finite systems may also be 
deduced from the $1/N$ expansion, which we consider in the following 
section. 

We will later compute the magnetization $M = \partial F / \partial 
B |_{\psi = 0}$, both in the thermodynamic limit and for systems of 
finite size, from the same type of free energy as 
in Eq.~(\ref{ess1l}). This undertaking is simplified in the light 
of insight gained from $1/N$ and RG analyses, and is thus deferred 
to Sec.~VI. At this point we note only that the magnetization 
calculated in this framework is not zero in the low-field regime, 
as would be required of a system with a spin gap and no broken 
symmetry. This is a consequence of the approximation, implicit 
in Eq.~(\ref{elde}) and in deriving the free energy $F(b)$ 
(Eq.~(\ref{ess1l})), that the field $b$ be large on the scale of 
$|\dot{\phi}|$. We may nevertheless conclude that analysis of the 
spin stiffness itself is qualitatively revealing of many properties 
of the system. 

\section{1/N Expansion}

To address the situation in the weak-field regime, we adopt instead 
a $1/N$ expansion, \cite{rp} which is expected to be appropriate in 
describing spin-gap phases. Here the staggered spin ${\bf n}$ is 
assumed to exist in an $N$-dimensional spin space, in which only 
the direction $n_z$ is selected by the magnetic field, and a 
controlled expansion may in principle be performed in the small 
quantity $1/N$ to compute fluctuation corrections to the 
saddle-point solution. The relevant parts of the action are ({\it 
cf.} Eq.~(\ref{esss})) 
\begin{equation}
   {\cal S}_E = \frac{1}{2 g} \int\! d\tau \, dx \,
   \left[
      \left( \partial_{\mu} {\bf n} \right)^2 -
      \bar{b}^2 \left( 1 - n_{z}^2 \right) -
      i \lambda \left( {\bf n}^2 - 1 \right)
   \right],
\label{es1n}
\end{equation}
where $\bar{b}$ denotes $b/c$, and the constraint that the spin 
${\bf n}$ have unit magnitude is made explicit with the Lagrange 
multiplier $i \lambda$. Integration over ${\bf n}$ and taking the 
functional derivative of the resultant effective action with 
respect to $\lambda$ yield an equation for the saddle-point value 
of the Lagrange multiplier, 
\begin{equation}
\frac{1}{g} = (N-1) \sum_{\bf k} \frac{1}{{\bf k}^2 + i \lambda} +
\sum_{\bf k} \frac{1}{{\bf k}^2 + i \lambda + \bar{b}^2},
\label{elc}
\end{equation}
in which the $B$-field term is found to appear only at $O(1/N)$. 
$i \lambda$ appears as a mass, or cutoff term in momentum 
integrations, and is thus an upper length scale for cooperative 
processes in the system, or simply a correlation length (inverse 
excitation gap). Thus by writing the saddle-point solution as $i 
\lambda = c^2 \pi^2 / \xi(B)^2$, and carrying out the summation at 
low $T$ from 0 to $\pi/a$, this correlation length is determined from 
\begin{eqnarray}
1 & = & - \frac{g}{2 \pi} \ln \left[ \left( \frac{a/\xi}{1 + \sqrt{1
+ (a/\xi)^2}} \right)^{N-1} \right. \label{e1nce} \nonumber \\ & &
\;\;\;\;\;\;\;\; \times \left. \left( \frac{\sqrt{(a/\xi)^2 + 
(a/L_m)^2}}{1 + \sqrt{1 + (a/\xi)^2 + (a/L_m)^2}} \right) \right] ,
\end{eqnarray}
which should be compared with $\rho_s = 0$ emerging from Eq.~(\ref{efsse}). 

In the weak-field regime, $\xi \ll L_m$, one finds 
\begin{equation}
\xi \simeq (1/2) a e^{2 \pi / N g} [ 1 + (a^2/8NL_{m}^2) e^{4 \pi / g}] .  
\label{exiwf}
\end{equation}
This is an $O(1/N)$ correction to the previous result, but there is 
also a difference of a power of $1/N$ in the exponent. Taking the 
field derivative, the result $\partial_b (c \pi / \xi)^2 = - 2 b / 
N$ ensures that the magnetization contribution from the ${\bf k}$
summation terms in $F$ are identically zero to $O(1/N)$.  At the
same order, the field derivative of the ${\bf k}$-independent term 
$b^2 + c^2 \pi^2 / \xi^2$ is $2b (1 - 1/N)$. The behavior required 
of a gapped system is obtained in the weak-field regime on making 
the well-known identification, deduced by comparison with exact
RG results (see Ref.\onlinecite{rp} and below) $N \rightarrow 
N - 2$. Returning to the physical situation of the 
$O(3)$ magnet, the magnetization is zero at all fields $B < B^*$. 
The saddle-point solution for $\xi (B)$ (\ref{e1nce}) becomes 
precisely that deduced from Eq.~(\ref{efsse}) (up to a negligible 
term retained from a lower cutoff), with the same field $B^*$ for 
divergence of $\xi$. For the $O(3)$ case we thus deduce the expected 
gapped state, with no breaking of spin rotational symmetry.

For completeness, we note the strong-field results of the $1/N$
treatment.  When the magnetic length is small ($L_m \ll \xi$), the
correlation length $\xi \simeq (1 / 2) a e^{2 \pi / g (N - 1)} (a /
L_m)^{1 / (N - 1)}$ has a direct field dependence. The magnetization
is given to leading order by $M \simeq - C + 2 b L_{m}^2 / [(N-1)
\xi^2]$. The quantity appearing in these expressions is not $N$ but
$N - 1$, indicating again the lowering of symmetry in the field
direction above a finite critical field $B^*$. However, these 
high-field results are not meaningful in the $O(3)$ case. 

In summary, the spin stiffness and $1/N$ techniques taken together 
yield a consistent picture at both weak and strong magnetic fields. 
However, while the Ansatz $N \rightarrow N - 2$ has been applied 
previously,\cite{rp} and leads to consistency between the results of 
this and the previous section, it cannot be taken to be entirely 
satisfactory. For this reason we proceed by applying the RG technique 
to the model of Eq.~(\ref{ese}). 

\section{Renormalization Group}

We consider next a renormalization-group (RG) study of the NLsM in 
an applied field. This approach yields meaningful results over the 
full parameter range. We adopt a standard Wilson momentum-shell 
treatment, further details of which are presented in App.~1. In 
brief, the action in the form
\begin{equation}
{\cal S}_E = \frac{1}{2 g} \int_{0}^{L_T} \!\!\!\!\! d \tau \!\! 
\int_{0}^{L} \!\!\!\! d x \, [ (\partial_{\mu} {\bf n})^2 - \bar{b}^2 
(1 - n_{z}^2) + 4 i \bar{b} n_x \dot{n}_y ]
\label{esrg}
\end{equation}
may be reexpressed in terms of variables $\phi$, representing
in-plane fluctuations of ${\bf n}$, and $\sqrt{g} \sigma_z = n_z$,
representing out-of-plane fluctuations. The latter is chosen to
facilitate a perturbative expansion in $g$, which yields
\begin{equation}
{\cal L}_E = {\mbox{$\frac{1}{2 g}$}} \left( A' - \bar{b}^2 \right) -
{\mbox{$\frac{1}{2}$}} \sigma_z ( - \partial_{\mu}^2 + \bar{b}^2 - A') 
\sigma_z + O(g).
\label{elrg}
\end{equation}
$A' = (\partial_{\mu} \phi)^2 + 2 i \bar{b} \dot{\phi}$ denotes
in-plane fluctuation terms, which, because $\phi(\tau,x)$ is assumed 
to vary slowly, can be taken to be a small constant (no fast Fourier 
modes) in the momentum shell $\gamma \Lambda < |{\bf k}| < \Lambda$
($\gamma \rightarrow 1$) around the finite upper cutoff $\Lambda = 
\pi / a$ set by the lattice length scale. By expanding in $A'$, the 
form of Eq.~(\ref{elrg}) is recovered, but with new coefficients 
$g(a')$ and $\bar{b}(a^{\prime})^2$ given at 1-loop order by
\begin{eqnarray}
\frac{1}{g (a^{\prime})} & = & \frac{1}{g_0} - {\rm Tr}^{\prime}
\frac{1}{- \partial_{\mu}^2 + \bar{b}^2}, \label{ergeif1} \\ \bar{b}
(a^{\prime})^2 & = & \bar{b}_{0}^2 - g_0 {\rm Tr}^{\prime} \ln 
\left( - \partial_{\mu}^2 + \bar{b}_{0}^2 \right). \label{ergeif2} 
\end{eqnarray}
The integrals represented by the partial traces (Tr$^{\prime}$) are 
performed over an isotropic momentum shell in 1+1d, with $\Lambda_T = 
\Lambda$. Alterations arising from considering instead a rectangular 
momentum shell, with separate integration over space and inverse 
temperature, would affect only non-universal prefactors. 

Inspection of the equation (\ref{ergeif1}) for renormalization of 
the coupling constant $g$ indicates already the effect of the field. 
Performing the momentum-shell integral yields 
\begin{equation}
\frac{g_0}{g(a^{\prime})} = 1 - \frac{g_0}{4 \pi} \ln \frac{ 1 + 
(a / L_m)^2}{(a / a^{\prime})^2 + (a / L_m)^2}
\label{ergcc}
\end{equation}
for $a \le a^{\prime} \le {\cal L} = {\rm min} [ L, L_T ]$, assuming 
(see below) that the flow of $L_m$ is at most weak. For weak fields 
$L_m \gg {\cal L}$ we see that $g_0 / g({\cal L}) = 1 - ( g_0 / 
2 \pi) \ln ({\cal L} / a)$, {\it i.e.} $g$ is renormalized by the 
full effects of quantum fluctuations. For strong fields ${\cal L} 
\gg L_m \gg a$, we have $g_0 / g({\cal L}) = 1 - ( g_0 / 2 \pi) 
\ln (L_m / a)$, and the coupling constant $g$ is independent of the 
system length scale ${\cal L}$. In this case renormalization is 
weak, and for very strong fields $L_m \ll a$ it vanishes ($g_0 / 
g({\cal L}) = 1$), demonstrating the suppression of quantum 
fluctuation effects by high fields. 

Following the standard RG procedure, we define the flow parameter 
$l = \ln (a^{\prime}/a)$, whence $dl = da / a$. Taking the derivative 
with respect to $l$ of Eqs. (\ref{ergeif1},\ref{ergeif2}), we obtain 
the differential form of the coupled RG equations  
\begin{eqnarray}
\frac{d g}{d l} & = & \frac{g^2}{2 \pi} \frac{1}{ 1 + \bar{\beta}^2},
\label{ergedf1} \\ \frac{d \bar{\beta}^2}{d l} & = & 2 \bar{\beta}^2
- \frac{g}{2 \pi} \ln \left( 1 + \bar{\beta}^2 \right).
\label{ergedf2}
\end{eqnarray}
These represent an extension of the usual RG equations to include a
magnetic field $B$, contained in $\bar{\beta} = a^{\prime} \bar{b}
(a^{\prime}) \; (= a^{\prime} b(a^{\prime}) / c)$. Eq.~(\ref{ergedf1}) 
is the conventional ``$\beta$-equation'', the terminology given to 
the renormalization of the coupling constant, but with an additional 
field term in the denominator. Clearly, a strong field restricts flow 
to the strong-coupling (disordered) limit, and, as observed in the 
previous paragraph, acts to suppress quantum fluctuation effects. 
This points to a ``deconfinement of excitations'' at suitably high 
field. Eq.~(\ref{ergedf2}) expresses the renormalization of the 
magnetic field term $b = {\tilde g} \mu_B B$ with dynamical exponent 
$z = 1$ (from the first term, which arises purely from the presence 
of $a^{\prime}$ in the quantity $\beta$), but with additional, 
logarithmic suppression of this flow at strong field and coupling 
(second term). 

At this point some commentary is in order regarding the above RG 
equations. The NLsM in the presence of a magnetic field may be 
expressed in the gauge theoretic form\cite{rf,rss}
\begin{equation}
{\cal L} = \frac{c}{2 g} \left[ \frac{1}{c^2} \left( \partial_{\tau} 
n_{\alpha} + \frac{i \tilde g \mu_B}{\hbar} \epsilon_{\alpha \beta 
\gamma} B_{\beta} n_{\gamma} \right)^2 + (\partial_x n_{\alpha})^2 
\right] ,
\label{egil}
\end{equation}
with the field term ${\tilde g} \mu_B B / c$ in the role of a gauge 
potential. From this it has been argued\cite{rss} that the field 
term should scale only with temperature (the $z = 1$ behavior of the 
first term in Eq.~(\ref{ergedf2})), and cannot be renormalized 
separately, for example by interactions, as this would violate the 
gauge invariance. This scaling hypothesis stands in contrast to the 
result (Eq.~(\ref{ergedf2})) of the Wilson momentum-shell RG approach 
taken above, where the second term indeed gives a field renormalization, 
albeit one which is logarithmically weak. However, the present situation 
is not unique: similar breakdown of this scaling occurs in other cases, 
particular examples being the Fermi liquid and the Luttinger 
liquid, also considered in Ref.\onlinecite{rss}, where there 
is a temperature-independent renormalization of the field 
term.\cite{footnote1,footnote2} We note also that on taking 
into account magnetic saturation (Sec.~II) in the present model, 
additional field terms would arise which preclude the use of a 
gauge-invariance argument. Finally, the qualitative features of 
the discussion are not affected by the logarithmic term in 
Eq.~(\ref{ergedf2}), because its field renormalization is weak. 

In analyzing the content of the RG equations 
(\ref{ergedf1},\ref{ergedf2}), we concentrate first on the fixed 
points, in order to obtain a qualitative picture of the RG flow 
diagram. Seeking a fixed point by a weak-field expansion around 
$\bar{\beta}_* = 0$, we find
\begin{equation}
\frac{dg}{dl} \simeq \frac{g^2}{2 \pi}, \;\;\;\;\;\; \frac{d \ln
\bar{\beta}^2}{dl} = 2 - \frac{d \ln g}{dl},
\end{equation}
which may be solved to yield
\begin{equation}
\frac{g_0}{g} = 1 - \frac{g_0}{2 \pi} l, \;\;\;\;\;\; \bar{\beta} =
\bar{\beta}_0 e^l \left( 1 - \frac{g_0}{2 \pi} l \right)^{1/2}.
\label{ergfp}
\end{equation}
The fixed point $(g_{*},\bar{\beta}_{*}) = (\infty,0)$ is evidently
stable if the flow is stopped at $l_* = 2 \pi / g_0$. The system
will flow to this strong-coupling regime if the starting value
$\bar{\beta}_0$ is sufficiently small. The length scale ${\cal L}_*
= a e^{l_*}$ at which the flow stops may be compared with the
spatial and thermal dimensions ($L,L_T$) of the system to calculate
directly the effects of finite size and temperature ({\it cf.} 
Sec.~III). At strong fields $(\bar{\beta} \rightarrow \infty)$, $dg/dl = 
0$, or $g = g_0$, and similarly $d \ln \bar{\beta}^2 = 2 dl$, from 
which it follows that $\bar{b}(l) = \bar{b}_0$, {\it i.e.} in this 
regime neither the coupling nor the field is renormalized.

\begin{figure}
\centerline{\psfig{figure=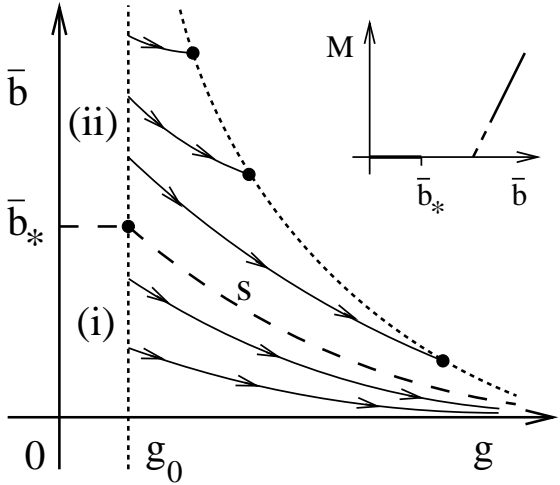,height=6.5cm,angle=270}}
\medskip
\caption{RG flow diagram for $g$ and $\bar{b}$. Strong- and
weak-coupling regimes are separated by separatrix $s$.}
\end{figure}

The RG equations (\ref{ergedf1},\ref{ergedf2}) may be solved 
numerically in the form 
\begin{equation}
\frac{\partial x}{\partial l} = \frac{-1}{l(1 + l^2 y^2)},  \;\;\;\; 
\frac{\partial y}{\partial l} = \frac{-1}{2xyl^3} \ln (1 + l^2 y^2), 
\label{ergns}
\end{equation}
where $x = 2 \pi / g$ is the inverse coupling constant and $y = a 
{\bar b}$ corresponds to the field variable $b / c$ only. Solution 
leads to the flow diagram in Fig. 4, which has the following 
interpretation. The regime (i) of weak initial $B$-field is the 
strong-coupling phase, with confinement of (gapped) excitations. 
Here, the assumption (underlying the perturbative RG treatment) of 
small $g$ becomes inconsistent, but one may 
still deduce the critical length scale ${\cal L}_*$, and that the
magnetization $M = 0$. In this region $O(3)$-symmetry is restored by
quantum fluctuations, which may thereby be considered as ``screening'' 
the magnetic field. In contrast, the regime (ii) of strong initial
$B$-field corresponds to weak coupling, where $g$ and $\bar{b}$ are
only weakly renormalized. Here, the excitations are deconfined on a
length scale ${\bar \xi(B)}$ whose flow is governed by $B$. In this
region, quantum fluctuations are suppressed by the magnetic field,
and the broken $O(3)$ symmetry cannot be restored.

The properties of the broken- and unbroken-symmetry phases may be
further contrasted by considering the correlation length $\xi$ in
each regime.  This is a physical quantity and so does not change
under the RG flow, whence $d \xi / d a^{\prime} = 0$ and
\begin{equation}
\frac{\partial g}{\partial l} \frac{\partial \xi}{\partial g} +
\frac{\partial \bar{\beta}^2}{\partial l} \frac{\partial
\xi}{\partial \bar{\beta}^2} + \xi = 0.
\label{exi}
\end{equation}
At low fields, we find that $- \partial \xi / \xi \simeq 2 \pi
\partial g / g$, which has the solution $\xi_0 = \xi e^{2 \pi (1/g_0
- 1/g)}$. If one assumes that $\xi \rightarrow a$ (the bare lattice
constant), as the system flows to the strong-coupling limit ($1/g_0
\rightarrow 0$), then $\xi_0 = a e^{2 \pi / g_0}$ returns the same,
finite physical correlation length as in the previous analyses. For
high values of the initial field we have $- 2 \partial \xi / \xi
\simeq \partial \bar{\beta}^2 / \bar{\beta}^2$, which leads to the
relation $\xi_0 = \xi {\cal L}/a$ ($\bar{b}_0$ is invariant) under
the RG flow. Thus, for any finite $\xi$ during the flow, the bare
correlation length $\xi_0$ is the system size $L$ or $L_T$,
corresponding to the quasi-long-range order expected of an XY system
in one spatial dimension.

Before concluding our RG analysis, we also consider its qualitative
predictions for the system magnetization, a measurable quantity.
This may be calculated as
\begin{equation}
M = \frac{\tilde{g} \mu_B a}{c} \frac{\partial F}{\partial 
\bar{\beta}_0} = \frac{\tilde{g} \mu_B a}{c} \left( \frac{\partial
\bar{\beta}}{\partial \bar{\beta}_0} \frac{\partial F}{\partial 
\bar{\beta}} + \frac{\partial g}{\partial \bar{\beta}_0} 
\frac{\partial F}{\partial g} \right).
\label{emrg}
\end{equation}
Once again, for weak fields we find that $\partial g / \partial
\bar{\beta}_0 = 0$ and $\partial \bar{\beta} / \partial
\bar{\beta}_0 = e^l \sqrt{1 - g_0 l / 2 \pi}$, where the latter
expression will introduce the characteristic length scale ${\cal L}$
of the short-range correlated phase. Now for any $F(\bar{b})$
analytic in $\bar{b}$,
\begin{equation}
M(B_0) \sim \sqrt{1 - \frac{g_0}{2 \pi} \ln \frac{\cal L}{a}},
\label{ergmlf}
\end{equation}
from which it is clear that the magnetization $M(B_0)$ vanishes in 
the scaling limit ${\cal L} \rightarrow {\cal L}^*$ corresponding 
to strong coupling ({\it i.e.} for all ${\bar b}_0$ sufficiently 
close to the fixed-point value ${\bar b}_0 = 0$. Within linear 
response, this vanishing of the magnetization $M = \chi_{\perp} B$ 
corresponds to the vanishing of the susceptibility $\chi_{\perp} 
\sim \ln (\xi / L)$ as the length of the system increases into the 
disordered state.\cite{ral} Hence the RG framework returns a zero 
magnetization in the regime where quantum fluctuations ensure no
spontaneous symmetry breaking. At high fields the magnetization is
finite, but its saturation in the physical system limits the useful
information which can be deduced from the above approach without 
inclusion of the saturation constraint (Sec.~II). 

\section{Magnetization}

We compute the magnetization from the Lagrangean ${\cal L}_{\rm E}$ 
of Eq.~(\ref{elde}), which has separate contributions from in- and 
out-of-plane spin fluctuations. The calculation is valid in the regime 
$|b| > \dot{\phi} \equiv \omega$ quoted in Sec.~II, and is thus 
effectively a high-field expansion. $\omega$ denotes the energy of 
excitations of the system, which may be taken to be spin waves 
($\omega = ck$) in the gapless regime. The magnetization is given by 
$M = {\tilde g} \mu_B \partial F / \partial b$, with $F = - \beta^{-1} 
\ln Z$ obtained from the partition function $Z = \int D \vartheta 
D \phi \exp (- S_{\rm E})$. Within the approximations specified 
below Eq.~(\ref{elde}), the separable nature of ${\cal L}_{\rm E}$ 
gives the result $M = M_o + M_i$. The out-of-plane contribution 
$M_o$ is purely dynamical, in the sense that $\vartheta$ 
fluctuations appear only in the form $(1 / 2 g) \vartheta \, G^{-1} 
\vartheta$. We may write $G^{-1} = G_{0}^{-1} - X$, with the 
bare propagator given by $G_{0}^{-1} = b^2 - \partial_{\tau}^2 - 
\partial_{x}^2$ and the corrections by $X = {\dot \phi}^2 + 2 i b 
{\dot \phi}$. In the high-field regime $|\dot{\phi} / b|$ is small, 
so that contributions from $X$ may be neglected. The appropriate 
piece of the partition function is readily evaluated as 
\begin{equation}
M_o = \frac{\beta}{2} \sum_{\bf k} \frac{1}{\sqrt{ b^2 + c^2 k^2}} 
\coth \left( {\textstyle \frac{1}{2}} \beta \sqrt{ b^2 + c^2 k^2} 
\right). 
\label{emdc}
\end{equation}
The assumption of an excitation spectrum linear up to the zone boundary 
is easily altered, for example to a sinusoidal form $\omega_k = |\sin 
ka|$ within the Brillouin zone, which gives rise only to small numerical 
corrections. Working consistently to lowest order in the small parameter 
$c/b$, and in limit of low $T$, we find that this contribution is simply 
a constant, $M_o = {\textstyle \frac{1}{2}} N_x \tilde{g} \mu_{B}$, 
which corresponds to a spatially uniform state. Small corrections 
of the form $M_{o}^{\prime} \sim B \ln B$ arise for either type of 
dispersion. 

Treatment of the in-plane variables $\phi$ at finite temperatures 
is more involved. We first represent the relevant part of the 
Lagrangean as 
\begin{equation}
{\tilde {\cal L}}_{\rm E} = \frac{1}{2 g} \left[ - b^2 - \phi ( 
\partial_{\tau}^2 + \partial_{x}^2 ) \phi + 2 i b {\dot \phi} \right] 
- \frac{i b}{\beta L} {\rm Tr} G_0 {\dot \phi} , 
\label{elp}
\end{equation}
to separate the term quadratic in $b$ which gives the classical, 
linear magnetization. A full treatment of $\phi \, \in \, [0,2\pi]$ 
involves decompactification to the interval $- \infty < {\tilde \phi} 
< \infty$ and restoration of periodicity by summation of the partition 
function over all winding number sectors $m \in {\cal Z}$. The phase 
is represented by $\phi (x, \tau) = {\tilde \phi} (x, \tau) + 2 \pi m 
\tau / \beta$, with ${\tilde \phi}$ a periodic function of the inverse 
temperature (${\tilde \phi} (x, \tau + \beta) = {\tilde \phi} 
(x, \tau)$). This form takes into account the possibility 
of windings in the space of $\tau$, whereas windings in $x$ may 
be safely neglected for a large system ($L \gg L_T$). On expanding 
the action, noting that all terms $\int_{0}^{\beta} d \tau {\dot
{\tilde \phi}}$ vanish (integral of total derivative), and 
discarding all terms independent of $b$ for the purposes of 
computing the magnetization, we obtain the corresponding partition 
function 
\begin{equation}
{\tilde Z} = C_0 \exp \left\{ - \frac{N_s b^2}{8 {\bar J}} \right\} 
\sum_{m = - \infty}^{+ \infty} \exp \left\{ \frac{N_s \pi^2}{2 {\bar 
J} \beta} m^2 + i 2 \pi \alpha m \right\} . 
\label{epfp}
\end{equation}
Here $C_0$ is a constant
and $\alpha = N_s b / 4 {\bar J} - N_x \ln (\epsilon + \sqrt{ 1 
+ \epsilon^2}) / 2 \epsilon$, with $\epsilon = \pi c / a b$, arises 
from the $\phi$-linear terms in Eq.~(\ref{elp}). Because $\epsilon$ 
is a small parameter in the approximation we employ, the second 
term in $\alpha$ simplifies to $N_x / 2$; this is also the origin of 
the results for $M_o$ and $M_{o}^{\prime}$. The sum over $m$ in 
Eq.~(\ref{epfp}) is simplified on recognizing that ${\tilde Z}$ may be 
represented by the $\theta_3$ function,\cite{ras} in terms of which 
the (logarithmic) $b$-derivative yields
\begin{equation}
M_i = - {\tilde g} \mu_B \left\{\frac{N_s b}{4 {\bar J}} + \frac{2 \pi 
N_s}{4 {\bar J} \beta} \sum_{m = 1}^{\infty} (-1)^m \frac{\sin (2 \pi 
\alpha m)}{\sinh (N_s \pi^2 m / 2 {\bar J} \beta)} \right\} .
\label{emfsc}
\end{equation}
The latter term is a sawtooth form, shown as the quantity $
M_f$ in Fig. 5(a), which is periodically extended from the 
interval $- 1/2 < \alpha < 1/2$ to all of $\alpha$. When superimposed 
on the linear part contained in the first term of Eq.~(\ref{emfsc}), 
$M_f$ gives the step-like form shown in Fig. 5(b). The width of the 
steps scales as $1/N_s$, so we see that $M_f$ is the finite-size 
correction to the linear magnetization. Such effects are of 
considerable interest in molecular magnets, where the small number 
of atoms gives rise to well-defined steps.\cite{rl} However, $M_f$ 
becomes indiscernible as the system size is taken to the thermodynamic 
limit, which is the condition we investigate below. In closing this 
section, we note that evaluation of the magnetization in the NLsM 
for weak fields $b$, and particularly around $B^*$, remains an open 
problem. Experimental comparison for this regime (Sec.~VIII) is 
facilitated by the knowledge (Secs.~III-V) that $M = 0$ below $B^*$.

\begin{figure}
\centerline{\psfig{figure=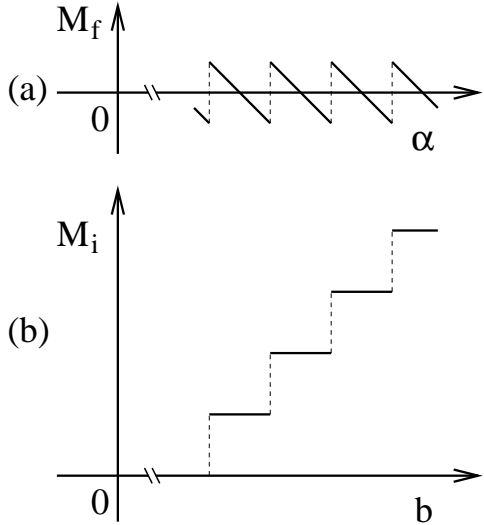,height=7.0cm,angle=0}}
\medskip
\caption{Schematic representation of finite-size magnetization 
contribution $M_i$ (a) to linear magnetization term (b) at  
high magnetic fields. }
\end{figure}

\section{Spin Correlation Functions}

Correlation functions describing the nature of the spin order may be 
calculated directly in the NLsM framework from 
\begin{eqnarray}
\langle {\hat T}_{\tau} {\hat S}_i (x,\tau) {\hat S}_j (0,0) \rangle 
& = & (-1)^x S^2 \langle n_i (x,\tau) n_j (0,0) \rangle \label{escf} 
\nonumber \\ & & \;\;\;\;\;\; + (aS)^2 \langle l_i (x,\tau) l_j (0,0) 
\rangle ,
\end{eqnarray}
and used to compare the properties of the system at low and high
fields. ${\hat T}_{\tau}$ is the ordering operator in imaginary time. 
In the low-field regime where full symmetry is retained, the 
gap $\Delta$ to all spin excitations results in exponentially decaying
spin correlations, reflecting the short-range nature of the quantum
disordered phase. 

Above the critical field, as discussed previously, the gap closes and 
the low-energy features are described by an XY model for the plane 
normal to the field. The correlation functions in this 
regime are discussed in detail in App. 2. Because the quantities 
$\langle n_z (x,\tau) n_z (0,0) \rangle$ and $\langle l_+ (x,\tau) 
l_- (0,0) \rangle$ ($l_{\pm}$ denotes $l_x \pm i l_y$) depend 
directly on the gapped, out-of-plane fluctuations in $\vartheta$, 
these remain short-ranged. However, $\langle l_z (x,\tau) l_z (0,0) 
\rangle$ is proportional to $\partial_{\tau}^2 \langle \phi(x,\tau) 
\phi(0,0) \rangle$, which leads to a power-law (or ``quasi-long-ranged'') 
decay of the correlation function parallel to the field direction,
\begin{equation}
\langle {\hat T}_{\tau} {\hat S}_z (x,\tau) {\hat S}_z (0,0) \rangle 
= \frac{\tilde B}{8 \pi g} \left[\frac{a^2}{(x - i \tau)^2} + 
\frac{a^2}{(x + i \tau)^2}\right] .
\label{escfpar}
\end{equation}
We have omitted here a constant term proportional to the square of 
the local magnetization. Similarly, $\langle n_+ (x,\tau) n_- (0,0) 
\rangle = \langle {\rm exp} ( i \phi (x, \tau) - i \phi (0,0)) 
\rangle$, which gives
\begin{equation}
\langle {\hat T}_{\tau} {\hat S}_+ (x,\tau) {\hat S}_- (0,0) \rangle = 
{\tilde C} (-1)^x \left( \frac{a^2}{x^2 + \tau^2} \right)^{(g / 4 \pi)} ,
\label{escfper}
\end{equation}
a separate power-law form for the correlation function perpendicular
to the field direction. ${\tilde B}$ and ${\tilde C}$ are constants. 
The results for the long-ranged parts of both longitudinal and 
transverse correlation functions are very similar to those of a 
Luttinger-liquid description of the gapless, broken-symmetry 
phase.\cite{rcg} The correlation exponent in Eq.~(\ref{escfper}) 
has a direct dependence on the bare coupling constant, and may be 
identified as an effective Luttinger-liquid exponent 
\begin{equation}
{\tilde K} = \frac{\pi}{2 g} = \frac{\pi S}{2} \sqrt{ \frac{J_x}{\bar J}} 
= 0.42.
\label{eelle}
\end{equation}
This quantity is independent of the field, and in the NLsM approach 
is rather close to the constant value\cite{rcg} ${\tilde K} = 1/2$ 
given by bosonization for all ladder systems.

\section{Comparison with Experiment}

We turn to a comparison between the preceeding physical ideas 
and experiments performed on the two-chain, $S = 1/2$ spin ladder
material CuHpCl. The magnetization\cite{rcclpmm} illustrates the
transition from the gapped regime of unbroken symmetry, and the
nature of the broken-symmetry phase, while nuclear magnetic resonance 
(NMR) measurements of spin relaxation rates\cite{rcfjhbhlp} reveal 
some characteristics of the spin correlation functions in both types 
of ground state.

From Sec.~VI, the magnetization takes the simple form 
\begin{equation}
M = - {\tilde g} \mu_B N_s (b / 4 \bar{J}) + M_o + O(1/N_y S) , 
\label{em}
\end{equation}
where $M_o$ (Eq.~(\ref{emdc})) becomes constant for the low temperatures 
of most interest. While the linear term is always present, we have 
shown in Secs.~IV and V that below a threshold field $B^*$, where 
the system has a spin gap, it is canceled by the corresponding 
correlation-length term. As discussed in Sec.~II, above an upper 
threshold ($B_{c2}$) the magnetization will saturate at the value 
$M_s$. Here we will for clarity treat saturation as a simple cutoff, 
rather than pursuing the more complex but fully systematic method 
of Sec.~II, which will be discussed elsewhere. Because the transition 
at $B_{c2}$ is from an XY phase with strongly suppressed quantum 
fluctuations to a fully polarized state with a spin-wave description, 
an exact treatment is of less interest in the current context. 
Finally, the $O(1/N_y S)$ term is included in Eq.~(\ref{em}) 
as a reminder that large corrections are to be expected in any 
quantitative comparison of a ``highly quantum'' (small $N_y S$) 
magnetic system with the semiclassical NLsM. 

Specializing to the parameters of CuHpCl, the exchange 
constants deduced from magnetization and susceptibility 
measurements\cite{rcclpmm,rhrbt} are $J^{\prime} / k_B$ = 13.2K 
and $J / k_B$ = 2.4K, whence $\bar{J} / \mu_B$ = 13.3T and $J_x 
/ \mu_B$ = 3.6T. Taking the simplest 
case of constant $M_o$, and the lower critical field $B_{c1}$ for 
onset where $M(B_{c1}) = 0$, we obtain $B_{c1} = \bar{J}/\tilde{g} 
\mu_B$ = 6.6T. The saturation field $B_{c2}$ is given from  $M(B_{c2}) 
= M_s$ as $B_{c2} = (4 S / \tilde{g} \mu_B) \bar{J}$ = 13.3T. These 
values agree well with a linear extrapolation of the magnetization 
data at the lowest temperature in Ref.\onlinecite{rcclpmm}, which 
yields $B_{c1}$ = 6.8T and $B_{c2}$ = 13.7T. The computed magnetization 
is shown in Fig. 6, where the dashed line indicates the validity limit
($|b| \gtrsim c \pi / a$) of the calculation.

\begin{figure}
\centerline{\psfig{figure=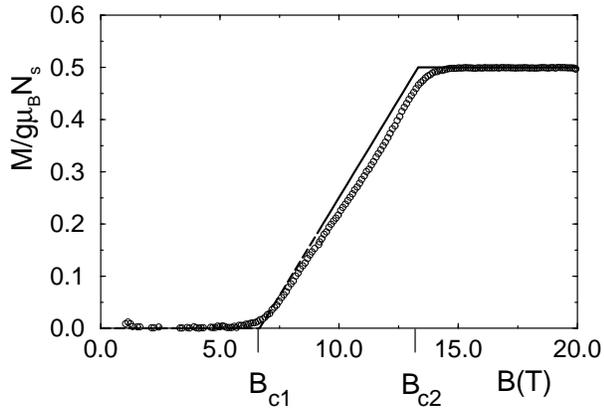,height=5.5cm,angle=270}}
\medskip
\caption{Computed magnetization for the spin ladder system CuHpCl 
(see text). Circles are data from Ref.~[5]. }
\end{figure}

The NLsM prediction of linear magnetization appears closer to the
data than a repulsive boson model\cite{ra,rhpl} and other purely 1D
approaches, such as bosonization\cite{rcg} and simulations on a single 
ladder\cite{rhpl} (or a Haldane chain,\cite{rkpst} which has the same 
universality class), all of which predict square-root cusps at 
$B_{c1}$ and $B_{c2}$ not evident in the data.\cite{rcclpmm} For 
a temperature $T$ = 0.42K $\sim \Delta / 25$ this form should 
be visible in the pure, 1d system \cite{rhpl}. This may be
indicative of a weak interladder coupling, which also causes the
real material to display 3D order at intermediate $B$ and low $T$.
The gradient of the linear region can be written as
\begin{eqnarray}
\chi_{{\rm NL} \sigma {\rm M}}^{\perp} & = & \frac{1}{N_s} \left( 
\frac{N_s S}{ 4 {\bar J} S - 2 {\bar J} / N_y} \right) \label{esan} 
\nonumber \\ & = & \chi_{\rm AF}^{\perp} \left( \frac{1}{1 - 1 / 
(2 N_y S)} \right),
\end{eqnarray}
where $\chi_{\rm AF}^{\perp} = (\tilde{g} \mu_B)^2 / 4 \bar{J} a^2$
is the N\'eel susceptibility per unit volume, the result for a
classical antiferromagnet with effective exchange coupling ${\bar J}$. 
The last term expresses the deviation from classical behavior as a 
function of spin magnitude and the number of coupled chains. That 
the magnetization adopts such a quasi-classical form in the 
broken-symmetry regime, at least in systems with no topological 
term, may be ascribed in part to suppression of quantum fluctuations 
by the field (effective beyond $B_{c1}$), and possibly also to the 
presence of higher-dimensional couplings.

Further commentary is in order concerning the agreement between the 
NLsM results and experiment. The semiclassical model has recently 
been used with considerable success, also for the field response, of 
small (spatially uniform) systems with rather small $S$.\cite{rcl} 
However, the apparent convergence of theory and experiment for the  
``maximally quantum'' case of $N_y S = 1$ in CuHpCl is in fact a 
somewhat fortuitous consequence of the parameter ratio $J / J_{\perp}$. 
The magnetization computed as above is valid for large $B$, and unlike 
the correlation length or gap is not sensitive to the coupling 
constants. Thus the model predictions for the isotropic ladder ($J / 
J^{\prime} = 1$), and for the Haldane chain (also $N_y S = 1$), are 
little different from the above, whereas the known results have much 
smaller gapped regions. Thus we stress again that the main value of 
the current approach lies in its qualitative features. From this and 
other work,\cite{rcl} it appears that $N_y S > 3$ is sufficient for 
the NLsM to provide quantitative accuracy, although the gap is then 
very small. 

The experimental comparison gives additional perspective on the 
utility of the spin stiffness and $1/N$ treatments (Secs.~III and IV). 
We deduce from $L_{m}^*$ that $B^* = 0.32\bar{J} / \tilde{g} \mu_B$ 
= 4.2T, a value rather lower than $B_{c1}$ above, and emphasize that 
in the regime between $B^*$ and $B_{c1}$ neither treatment is reliable. 
Following the RG analysis, the field scale $\bar{b}_*$ from the 
numerical solution gives $B_*$ = 1.6T, although prefactor corrections 
(above) can be expected. 

The other category of experiments performed on CuHpCl is the 
measurement of NMR spin relaxation rates, which probe the spin-spin 
correlation functions discussed in Sec.~VII. The experimental 
observable is the spin-lattice relaxation rate, given in $d$ 
dimensions by 
\begin{equation}
\frac{1}{\rm T_1} = \lim_{\omega \rightarrow 0} \frac{2 k_{\rm B}
T}{ \hbar \omega} \sum_{ij} \int \frac{d^d q}{(2 \pi)^d} F_{ij} ({\bf q}) 
\chi_{ij} ({\bf q},\omega,T) ,
\label{enmrslrr}
\end{equation}
in which $F_{ij}$ are hyperfine coupling constants. This is obtained 
from the dynamic susceptibility
\begin{eqnarray}
   \chi_{ij} ({\bf q},\omega,T) &=& - i \int\! dt \, d {\bf x} \,
   e^{i \omega t + i {\bf q} \cdot {\bf x}}
   \nonumber \\ && \mbox{} \times \Theta(t)
   \left\langle
      \left[ {\hat S}_i ({\bf x},t) , {\hat S}_j ({\bf 0},0) \right]
   \right\rangle_T , 
\label{eds}
\end{eqnarray}
which in turn makes contact with the results of Sec.~VII. By 
straightforward scaling, spin-spin correlation functions $\left\langle
{\hat T}_{\tau} {\hat S}_i (x,\tau) {\hat S}_j (0,0) \right\rangle$ 
decaying exponentially in 1+1d space correspond to relaxation rates 
which are exponentially activated in temperature, for $T$ below the 
characteristic energy scale $\Delta$. This is observed in experiments 
for fields $B < B_{c1}$.\cite{rcfjhbhlp} Similarly, power-law spatial 
decays correspond to thermal power laws in relaxation rates: by 
performing separate Fourier transformations on $x - i \tau$ and $x + 
i \tau$ and extracting the energy-dependence in the long-wavelength 
limit ($k \rightarrow 0$), if 
\begin{equation}
\langle {\hat T}_{\tau} {\hat S}_i (x,\tau) {\hat S}_j (0,0) 
\rangle \propto \delta_{ij} (x^2 + \tau^2)^{-\alpha} 
\label{enmrss1}
\end{equation}
then
\begin{equation}
1 / {\rm T_1} \propto T^{(2 \alpha - 2) + 1} ,
\label{enmrss2}
\end{equation} 
whence from Eqs. (\ref{escfpar},\ref{escfper}) the relaxation rate 
for the ladder system is 
\begin{equation}
\left. \frac{1}{\rm T_1} \right|_{\rm ladder} = A_{\rm \parallel} T + 
B_{\rm \perp} T^{- g / 2 \pi - 1} . 
\label{enmrss3}
\end{equation}
$A_{\rm \parallel}$ and $B_{\rm \perp}$ are constants related 
to the hyperfine coupling. An identical result was obtained in the 
Luttinger-Liquid formulation of Ref.\onlinecite{rcg}. In addition to 
the activated form in the gapped regimes at $B < B_{c1}$ and $B > B_{c2}$, 
the experimental measurements of $1/T_1$\cite{rcfjhbhlp} show clear
power-law divergence in the intermediate field regime, fully consistent
with expectations (Sec.~VII) based on a model of symmetry breaking by 
the effect of the field on quantum fluctuations. However, because the 
data\cite{rcfjhbhlp} cover at most one decade in temperature, while 
typically 3-4 decades are required to identify the exponent within 
meaningful bounds, we may not attempt to make any quantitative 
deductions (Eq.~(\ref{eelle})) concerning the coupling constant $g$. 

\section{Summary}

In conclusion, the nonlinear $\sigma$ model description of quantum
antiferromagnets in an external field provides a versatile framework
for illustrating the breaking and restoration of symmetries. The
case of $O(3)$ spins has a variety of physical realizations, which
illustrate clearly the transition from the gapped, short-range
ordered phase where full spin symmetry is maintained by quantum
fluctuations, to a gapless, quasi-long-range ordered regime where
the $O(3)$ symmetry is reduced to $O(2)$ by the field.

The foregoing analysis is not restricted to CuHpCl, but applies to
any system with a trivial topological term $\sum_i P_i$ 
(Eq.~(\ref{eswz})). We comment here on its 
application to materials in two other classes. CFTD \cite{rhcmmv} 
is a planar $S = {\textstyle \frac{1}{2}}$ Heisenberg AF, of a type 
studied extensively in connection with the high-temperature 
superconductivity problem,\cite{rchn} and known \cite{rem} to have AF 
order at $T = 0$. In an applied field one expects linear magnetization 
characteristics with the N\'eel susceptibility, beginning at $B = 0$, 
but not saturating because the superexchange interaction $J$ = 9.4meV 
is beyond the reach of laboratory fields. Magnetization results on 
further recently-synthesized, planar Cu compounds with significantly 
smaller exchange constants\cite{rawlt} may show evidence of a logarithmic 
approach to saturation, as expected from spin-wave theory for a purely 
2d system, and obtainable in the NLsM by inclusion of the total spin 
contraint as in Sec.~II. The cases of most interest are those exhibiting 
the physics of the Haldane gap. The primary example is the $S = 1$ AF
chain, of which NENP is considered a prototypical case (but for the
complication of a large single-ion anisotropy). The present study
yields the same qualitative features of a gapped regime with zero
magnetization, followed by approximately linear behavior towards
saturation (which could not be achieved), as in experiment.\cite{ragksi} 
Quantitative accord with experimental transition fields, and numerical 
simulations,\cite{rkpst} does not match that of Sec.~VIII for CuHpCl. 
Finally, there has been considerable recent interest in the possibility 
of field-induced magnetization plateaus in certain systems: in a NLsM
approach these may be expected, for example in $S > 1$ chains, when
the field strength is such that the projected in-plane spin $|S {\bf
n}_{\perp}|$ is of integer amplitude, leading to a gapped phase.

\section{Acknowledgments}

   We are grateful to G. Chaboussant, F. Mila, and S. Sachdev for 
useful discussions, and thank the Swiss National Fund for financial 
support. BN acknowledges the generosity of the Treubelfonds.

\appendix
\section{Renormalization group equations}

The appendix presents the derivation of the integral forms 
(\ref{ergeif1},\ref{ergeif2}) of the RG equations for the coupling 
constant and field. We begin with the Lagrangean for the NLsM in a 
magnetic field in the form of Eq.~(\ref{esrg}), 
\begin{equation}
{\cal L}_E = \frac{1}{2 g} [ (\partial_{\mu} {\bf n})^2 - \bar{b}^2 
(1 - n_{z}^2) + 4 i \bar{b} n_x \dot{n}_y ] ,
\label{elrga}
\end{equation}
where the parameters are defined in Secs.~II and III, and the last 
term is obtained by partial integration of one component of the cross 
product. The staggered magnetization satisfies $|{\bf n}|^2 = 1$ and 
(Sec.~II) lies predominantly in the ($x,y$) plane, so that we may choose 
$|n_z| \ll 1$ and specify the components of ${\bf n}$ by ($n_x = 
\sqrt{1 - n_{z}^2} \cos \phi$, $n_y = \sqrt{1 - n_{z}^2} \sin \phi$, 
$n_z$), with $0 \le \phi < 2 \pi$. Using these variables, the cross 
product term gives $n_x {\dot n}_y - n_y {\dot n}_x = (1 - n_{z}^2) 
{\dot \phi}$, and the gradient term becomes
\begin{equation}
(\partial_{\mu} {\bf n})^2 = (\partial_{\mu} n_z)^2 + (1 - n_{z}^2)
(\partial_{\mu} \phi)^2 + \frac{(n_z \partial_{\mu} n_z)^2}{1 - n_{z}^2} .
\label{egtrg}
\end{equation}
On making the substitution $n_z = \sqrt{g} \sigma_z$, where the coupling 
constant $g$ is taken to be a small parameter (weak-coupling expansion), 
the Lagrangean (\ref{elrga}) is exactly represented as 
\begin{eqnarray}
{\cal L}_E & = & {\mbox{$\frac{1}{2 g}$}} (1 - g \sigma_{z}^2) 
[(\partial_{\mu} \phi)^2 + 2 i {\bar b} {\dot \phi} - {\bar b}^2] + 
{\mbox{$\frac{1}{2}$}} (\partial_{\mu} \sigma_z)^2 \label{elrge} 
\nonumber \\ & & \;\;\;\; + \frac{1}{2} \frac{g}{(1 - g \sigma_{z}^2)} 
(\sigma_z \partial_{\mu} \sigma_z)^2 .
\end{eqnarray}
With the definition $A' = (\partial_{\mu} \phi)^2 + 2 i \bar{b} 
\dot{\phi}$, ${\cal L}_E$ may be expanded in the power series in $g$ 
\begin{eqnarray}
{\cal L}_E & = & {\mbox{$\frac{1}{2 g}$}} [A' - {\bar b}^2] + 
{\mbox{$\frac{1}{2}$}} (\partial_{\mu} \sigma_z)^2 - {\mbox{$\frac{1}
{2}$}} (\sigma_z)^2 [A' - {\bar b}^2] \label{elrgps} \nonumber \\ 
& & \;\;\;\; - {\mbox{$\frac{1}{2}$}} g (\sigma_z \partial_{\mu} 
\sigma_z)^2 + {\mbox{$\frac{1}{2}$}} g^2 \sigma_{z}^2 (\sigma_z 
\partial_{\mu} \sigma_z)^2 + O(g^3) .
\end{eqnarray}

We proceed to one-loop order by neglecting those terms $O(g)$ and higher, 
and by following the Wilson RG technique of integrating out states of 
higher momenta around a finite upper cutoff.\cite{rjk,ref,raa} In this 
process the renormalizing effect of the fast Fourier modes of $\sigma_z$ 
is taken into account, whereas the slow variable $\phi(\tau,x)$ and its 
(small) derivative terms give no contribution at high momenta. We compute 
the quantity 
\begin{equation}
I = \int {\cal D} \sigma_z ({\bf k}) e^{-{\mbox{$\frac{1}{2}$}} \int 
d {\bf x} \left[ {\mbox{$\frac{1}{2 g}$}} (A' - {\bar b}^2) + \sigma_z 
(- \partial_{\mu}^2 - A' + {\bar b}^2) \sigma_z + O(g) \right] } 
\label{ergi}
\end{equation}
in the isotropic momentum shell $\gamma \Lambda < |{\bf k}| < \Lambda$,
where $\gamma < 1$ and $\Lambda = \pi / a$. Evaluation of the Gaussian 
integral yields
\begin{eqnarray}
I & = & \prod_{\gamma \Lambda < |{\bf k}| < \Lambda} \sqrt{\frac{2 \pi}
{{\bf k}^2 - A' + {\bar b}^2}} \, e^{-\mbox{$\frac{1}{2 g}$} [A' - 
{\bar b}^2]} \label{ergie} \nonumber \\ & = & e^{-{\mbox{$\frac{1}{2}$}} 
{\rm Tr}^{\prime} \ln (- \partial_{\mu}^2 - A' + {\bar b}^2 )} \, 
e^{-\mbox{$\frac{1}{2 g}$} [A' - {\bar b}^2]} ,  
\end{eqnarray}
in which the small quantity $A'$ permits the expansions
\begin{eqnarray}
{\rm Tr}^{\prime} \ln (- \partial_{\mu}^2 - A' + {\bar b}^2 ) 
& = & {\rm Tr}^{\prime} \ln (- \partial_{\mu}^2 + {\bar b}^2 ) 
\label{ergte} \\ & & - {\rm Tr}^{\prime} \left[ \frac{1}{- 
\partial_{\mu}^2 + {\bar b}^2} A' \right] + O(A')^2 \nonumber
\end{eqnarray}
and 
\begin{equation}
{\rm Tr}^{\prime} \left[ \frac{1}{- \partial_{\mu}^2 + {\bar b}^2} 
A' \right] \simeq {\rm Tr}^{\prime} \left[ \frac{1}{- \partial_{\mu}^2 
+ {\bar b}^2} \right] A' .  
\label{ergtf}
\end{equation}
Substituting these results into the original Lagrangean ${\cal L}_E$ 
(\ref{elrgps}) returns the new, effective Lagrangean on renormalized 
lattice length scale $a'$,  
\begin{eqnarray}
{\cal L}_{\rm eff} & = & \frac{1}{2} \left( \frac{1}{g_0} - {\rm Tr}' 
\frac{1}{- \partial_{\mu}^2 + {\bar b}^2} \right) A' \label{eleffps} 
\nonumber \\ & & \;\;\;\; + \frac{1}{2} (\partial_{\mu} \sigma_z)^2 
- \frac{1}{2} (\sigma_z)^2 [A' - {\bar b}^2] \nonumber \\ & & \;\;\;\; 
- \frac{1}{2} \left( \frac{1}{g_0} {\bar b}^2 - {\rm Tr}' \ln (- 
\partial_{\mu}^2 + {\bar b}^2) \right) . 
\end{eqnarray}
Finally, by comparison of ${\cal L}_{\rm eff}$ with ${\cal L}_E$, we 
may identify the renormalized coupling and magnetic field terms as 
expressed in Eqs. (\ref{ergeif1},\ref{ergeif2}). 

\section{Spin correlation functions}

In this appendix we present the calculation of spin-spin correlation 
functions, in real space and inverse temperature, for the gapless, 
high-field phase of the NLsM. The latter part of Sec.~VII illustrates 
the transformation of these quantities for comparison with experiment. 
It is apparent from Eq.~(\ref{escf}) that four such functions must be 
considered, corresponding to correlations of the staggered and uniform 
spin components in directions parallel ($i,j = z$) and perpendicular 
($i,j = \pm$) to the field. We reexpress the spin components ${\bf n}$ 
and ${\bf l}$ in terms of the angles $\vartheta$ and $\phi$ (Sec.~II), 
for which the propagators are known. 

The parallel response function for the staggered spin components is 
computed using 
\begin{equation}
\langle n_z ({\bf x}) n_z ({\bf 0}) \rangle = \int {\cal D} {\bf n} 
n_z ({\bf x}) n_z ({\bf 0}) e^{- {\cal S} [{\bf n}]} ,
\label{ecfspa}
\end{equation} 
where $n_z = \cos \theta = \sin \vartheta \simeq \vartheta$ for small 
angular deviations from the plane determined by the field. One expects 
the linear expansion to be valid for the lowest-order form of the 
correlation function, and may use it also in the approximation $\int 
{\cal D} (\cos \vartheta) \simeq \int {\cal D} \vartheta$. Now 
\begin{eqnarray}
\langle n_z (x,\tau) n_z (0,0) \rangle = &&\int {\cal D} \vartheta \, 
\vartheta ({\bf x}) \vartheta ({\bf 0}) {\rm exp} \left\{ - {\mbox{$
\frac{1}{2 g}$}} \vartheta \cdot G_{0}^{-1} \vartheta \right\} 
\label{escfpa1} \nonumber \\ = \lim_{j \rightarrow 0} \frac{\delta}
{\delta j_x} \frac{\delta}{\delta j_0} \int {\cal D} \vartheta && {\rm 
exp} \left\{ - {\mbox{$\frac{1}{2 g}$}} \vartheta \cdot G_{0}^{-1} 
\vartheta  + j \cdot \vartheta \right\} , 
\end{eqnarray} 
in which $\vartheta$ denotes a vector representing states at all ${\bf 
x} = (\tau, x)$, $G_{0}^{-1} = - \partial_{\mu}^2 + {\bar b}^2$ is the 
matrix propagator for modes of $\vartheta$, and the current $j ({\bf 
x})$ is taken to zero at the end of the calculation. Evaluation of the 
Gaussian integral yields 
\begin{equation}
\lim_{j \rightarrow 0} \frac{\delta}{\delta j_x} \frac{\delta}{\delta 
j_0} {\rm exp} \left\{ {\mbox{$\frac{1}{2}$}} g \, j \cdot G_{0} j 
\right\} = g G_0 ({\bf x},{\bf 0}) , 
\label{espa2}
\end{equation} 
from which the propagator component is given by resolution into plane 
wave states according to 
\begin{equation}
G_0 ({\bf x},{\bf 0}) = \langle {\bf x} | \frac{1}{ - \partial_{\mu}^2 
+ {\bar b}^2} | {\bf 0} \rangle = \frac{1}{L L_T} \sum_{\bf k} \frac{
e^{i {\bf k \cdot x}}}{{\bf k}^2 + {\bar b}^2} , 
\label{espa3}
\end{equation} 
where ${\bf k} = (\omega_n, k)$. For the interesting limit of large 
distance ($ {\bar b} \sqrt{ x^2 + {\tau}^2} > 1$), we conclude that 
the correlation function
\begin{equation}
\langle n_z (x,\tau) n_z (0,0) \rangle = \frac{\tilde A}{\sqrt{{\bar 
b} \sqrt{ x^2 + {\tau}^2}}} \, e^{-{\bar b} \sqrt{ x^2 + {\tau}^2}} , 
\label{ecfspaf}
\end{equation} 
where ${\tilde A}$ is a constant, has the exponentially decaying form 
in space and inverse temperature expected from its origin in the 
gapped, out-of-plane excitations. 

The other component of the parallel response function is\cite{ral} 
\begin{eqnarray}
\langle l_z ({\bf x}) l_z ({\bf 0}) \rangle & = & \langle l_{z}^{\rm 
cl} ({\bf z}) l_{-}^{\rm cl} ({\bf 0}) \rangle \label{ecfupa} \nonumber 
\\ & & + \frac{g}{4 c} \frac{1}{L L_T} \sum_{\bf q} e^{i {\bf q \cdot 
x}} \left( 1 - \langle n_z ({\bf x}) n_z ({\bf 0}) \rangle \right) .
\end{eqnarray}
The latter term is a quantum correction arising from the constraint 
${\bf l} \cdot {\bf n} = 0$, which is unaffected by the field, and 
has the value $(g / 4 c)( \delta_{{\bf x},{\bf 0}} - \langle (n_z 
({\bf 0}))^2 \rangle )$, {\it i.e.} is a local quantity which 
contributes only to a constant proportional to the square of the 
local magnetization. From Eq.~(\ref{elcl}), the expression for 
$l_{z}^{\rm cl}$ may be simplified by noting that the field term 
${\bf b} = (0,0,b_z)$ selects $n_z$, which in turn introduces 
$\vartheta$, and thus only exponentially decaying contributions. 
The remaining term is $({\bf n} \wedge {\dot {\bf n}})_z = {\dot 
\phi} \sin^2 \vartheta \simeq {\dot \phi}$ only. With the exception 
of further constant terms ($\langle n_{z}^2 ({\bf 0}) \rangle$), 
which are neglected both here and in Eq.~(\ref{escfpar}), we may 
restrict our considerations to 
\begin{eqnarray}
(a S)^2 \langle l_z ({\bf x}) l_z ({\bf 0}) \rangle & = & - \frac{1}
{(4 {\bar J})^2} \, \langle {\dot \phi} ({\bf x}) {\dot \phi} ({\bf 0}) 
\rangle \label{eupa1} \nonumber \\ & = & \frac{1}{(4 {\bar J})^2} \,  
\partial_{\tau}^2 \langle \phi ({\bf x}) \phi ({\bf 0}) \rangle . 
\end{eqnarray}
We see now that we are dealing only with the gapless, in-plane modes, 
which can be expected to give a power-law correlation function at 
long distances, and the treatment is identical to the Luttinger liquid. 
By the same technique as above (Eqs. (\ref{escfpa1},\ref{espa2})), the 
correlation function for the azimuthal angles is given by 
\begin{equation}
\langle \phi ({\bf x}) \phi ({\bf 0}) \rangle = 2 g {\tilde B} 
{\tilde G}_0 ({\bf x}, {\bf 0}) ,
\label{eaacf}
\end{equation} 
with propagator ${\tilde G}_0 = - \partial_{\mu}^2$ for in-plane modes, 
and ${\tilde B}$ a constant. Use of the result (expressed in rectangular 
coordinates with $c$ retained explicitly)
\begin{eqnarray}
\frac{c}{L L_T} \sum_{\omega, k \ne 0} \frac{\omega^2}{k^2 + (\omega/c)^2} 
\, e^{i kx + i \omega \tau} & = & \frac{c^3}{2 \pi} \frac{x^2 - c^2 \tau^2}
{[x^2 + c^2 \tau^2]^2} \label{eddcfr} \nonumber \\ \;\; = \frac{c^3}{4 
\pi} \left[ \frac{1}{(x - i c \tau)^2} \right. & + & \left. \frac{1}{(x 
+ i c \tau)^2} \right]  
\end{eqnarray} 
leads directly to Eq.~(\ref{escfpar}). The prefactor is obtained from 
the identity $c / 2 {\bar J} = a / g$, while further identification of 
$\pi / 2 g$ with ${\tilde K}$ (\ref{eelle}) yields the result of 
Ref.\onlinecite{rcg}. Because we are not interested in finite-size 
properties for discussing the asymptotic behavior of correlation 
functions, we ignore the zero modes (Eq. (\ref{eddcfr})) occurring in 
the Luttinger liquid. Their presence is connected with decompactification 
of $\phi$ and the associated winding number (Sec.~VI), and their treatment 
discussed in Ref.~\onlinecite{rdl}. 

Moving to the correlation functions normal to the applied field, that 
for the staggered spin components has the simple form 
\begin{equation}
\Xi = \langle n_+ ({\bf x}) n_- ({\bf 0}) \rangle = \langle {\rm 
exp} ( i \phi ({\bf x}) - i \phi ({\bf 0})) \rangle .
\label{ecfspe}
\end{equation}
Defining the function $j({\bf x}') = \delta ({\bf x} - {\bf x}') - 
\delta({\bf x}')$, and using the equalities ${\tilde G}_0 ({\bf x}, 
{\bf x}) = {\tilde G}_0 ({\bf 0}, {\bf 0})$ and ${\tilde G}_0 ({\bf x}, 
{\bf 0}) = {\tilde G}_0 ({\bf 0}, {\bf x})$, permits one to write 
\begin{eqnarray}
\Xi & = & \int {\cal D} \phi \exp \left\{ - {\mbox{$\frac{1}{2 g}$}} 
\phi \cdot {\tilde G}_{0}^{-1} \phi  + i j \cdot \phi \right\} 
\label{espe1} \nonumber \\ & = & ({\rm Det} {\tilde G}_{0}^{-1})^{-1/2} 
\exp \left\{ - {\mbox{$\frac{1}{2}$}} g j \cdot {\tilde G}_{0} j 
\right\} \nonumber \\ & = & C' \exp \left\{ - g [{\tilde G}_0 ({\bf 0}, 
{\bf 0}) - {\tilde G}_0 ({\bf x}, {\bf 0})] \right\} ,
\end{eqnarray}
where $C'$ is a constant. The propagators are evaluated as above, to 
give 
\begin{equation}
{\tilde G}_0 ({\bf 0}, {\bf 0}) = \frac{1}{L L_T} \sum_{{\bf k} \ne 
{\bf 0}} \frac{1}{{\bf k}^2} = \frac{1}{2 \pi} \ln \left( \frac{{\cal 
L}}{a} \right) ,  
\label{ecfsp0}
\end{equation}
and 
\begin{equation}
{\tilde G}_0 ({\bf x}, {\bf 0}) = \frac{1}{L L_T} \sum_{{\bf k} \ne 
{\bf 0}} \frac{e^{i {\bf k} \cdot {\bf x}}}{{\bf k}^2} = - \frac{1}{4 
\pi} \ln \left( \frac{(2 \pi)^2 (x^2 + {\tau}^2)}{{\cal L}^2} \right) ,  
\label{ecfspx}
\end{equation}
both of which are cut off by the system length scale (Sec.~V) ${\cal 
L} = {\rm min} [L, L_T]$. Finally, 
\begin{equation}
\langle n_+ ({\bf x}) n_- ({\bf 0}) \rangle = {\tilde C} \left( 
\frac{{\cal L}}{a} \right)^{-\mbox{$\frac{g}{2 \pi}$}} \left( 
\frac{x^2 + {\tau}^2}{{\cal L}^2} \right)^{-\mbox{$\frac{g}{4 
\pi}$}} ,
\label{ecfspf}
\end{equation}
for altered constant ${\tilde C}$, returns the result in 
Eq.~(\ref{escfper}). 
That this is the only power-law contribution to $\langle S_+ ({\bf 
x}) S_- ({\bf 0}) \rangle$ is easily seen from 
\begin{eqnarray}
\langle l_{+}^{\rm cl} ({\bf x}) l_{-}^{\rm cl} ({\bf 0}) \rangle & = 
& \langle e^{i \phi ({\bf x}) - i \phi ({\bf 0})} {\dot \vartheta} 
({\bf x}) {\dot \vartheta} ({\bf 0}) \rangle \label{eucfper} \nonumber 
\\ & = & - \langle e^{i \phi ({\bf x}) - i \phi ({\bf 0})} \rangle [ 
\partial_{\tau}^2 \langle \vartheta ({\bf x}) \vartheta ({\bf 0}) \rangle ] , 
\end{eqnarray}
whose dependence on $\vartheta$ guarantees an exponential decay of 
correlations.


\begin{thebibliography}{99}
\bibitem{gf} We have made the assumption, well justified for the 
parameters of CuHpCl (Sec.~VIII), that all relevant energies should 
be below the finite-size gap, $E_0 = 2 \pi S J_y$, of spin waves 
transverse to the ladder.
\bibitem{footnote2}
We note that the gauge theory (Eq.~(\ref{egil})) is non-Abelian,\cite{rss} 
an important consequence of which is that different field components 
need not renormalize in the same way. Thus the result of 
Eq.~(\ref{ergedf2}) does not conflicit with the discussion of 
gauge invariance and RG methods introducing finite momentum cutoffs 
in Refs.\onlinecite{rhlm,rcln}, which concerns the Abelian case.
\end{thebibliography}
\end{document}